\documentclass[aps,prl,twocolumn,superscriptaddress,nofootinbib,floatfix]{revtex4-1}

\usepackage{graphicx}
\usepackage{amsmath}
\usepackage{amssymb}
\usepackage{color}
\usepackage{mathtools}
\usepackage{eufrak}
\usepackage[justification=RaggedRight]{caption}
\usepackage[breaklinks,colorlinks,citecolor=blue,linkcolor=red]{hyperref}
\usepackage[all]{hypcap}

\DeclareMathOperator{\sech}{sech}

\DeclareMathOperator{\arctanh}{arctanh}
\DeclareMathOperator{\atanhopt}{arctan[h]}

\newcommand{\beq}{\begin{equation}}
\newcommand{\eeq}{\end{equation}}
\newcommand{\bea}{\begin{eqnarray}}
\newcommand{\eea}{\end{eqnarray}}
\newcommand{\tpeak}{t_{\rm p}}
\newcommand{\tstart}{t_{\rm 0}}
\newcommand{\Apeak}{A_{\rm p}}
\newcommand{\rlr}{r_{\rm lr}}
\def\leq{\raise 0.4ex\hbox{$<$}\kern -0.8em\lower 0.62ex\hbox{$-$}}
\def\geq{\raise 0.4ex\hbox{$>$}\kern -0.7em\lower 0.62ex\hbox{$-$}}
\def\lsim{\raise 0.4ex\hbox{$<$}\kern -0.8em\lower 0.62ex\hbox{$\sim$}}
\def\gsim{\raise 0.4ex\hbox{$>$}\kern -0.7em\lower 0.62ex\hbox{$\sim$}}
\def\pm{\,\raise 0.4ex\hbox{$+$}\kern -0.8em\lower 0.62ex\hbox{$-$}\,}
\def\mp{\,\raise 0.4ex\hbox{$-$}\kern -0.8em\lower 0.62ex\hbox{$+$}\,}

\begin{document}

\title{Analytical Black-Hole Binary Merger Waveforms}
\author{Sean T. McWilliams}
\email{Sean.McWilliams@mail.wvu.edu}
\affiliation{Department of Physics and Astronomy, West Virginia University, Morgantown, West Virginia 26506, USA}
\affiliation{Center for Gravitational Waves and Cosmology, West Virginia University, Chestnut Ridge Research Building, Morgantown, West Virginia 26505, USA}

\date{\today}

\begin{abstract}

We present a highly accurate, fully analytical model for the late inspiral, merger, 
and ringdown of black-hole binaries with arbitrary mass ratios and spin vectors and the associated
gravitational radiation,
including the contributions of harmonics beyond the fundamental mode.  
This model assumes only that nonlinear effects remain small throughout the entire coalescence, and is developed
based on a physical understanding of the dynamics of late stage binary evolution, in particular on the 
tendency of the dynamical binary spacetime to behave like a linear perturbation of the stationary merger-remnant spacetime, even
at times before the merger has occurred.
We demonstrate that our model agrees with the most accurate numerical relativity results to within their own 
uncertainties throughout the merger-ringdown phase, and it does so for example cases spanning
the full range of binary parameter space that is currently testable with 
numerical relativity. 
Furthermore, our model maintains accuracy back to the innermost stable
circular orbit of the merger-remnant spacetime over much of the relevant parameter space, greatly
decreasing the need to introduce phenomenological degrees of freedom to describe the late inspiral.
 
\end{abstract}

\maketitle

\paragraph{Introduction.---}
\label{sec:intro}
Prior to the wide-ranging successes of numerical relativity (NR) that began with technical
breakthroughs in 2005 \cite{Pretorius:2005gq,Campanelli:2005dd,Baker:2005vv} (see \cite{NRreview} for
a recent review), the challenge of calculating
the gravitational-wave emission from a pair of merging black holes was seen primarily as a problem on the boundary
of nonlinear mathematics and computer science. The nonlinear nature of the partial differential equations
describing general relativity was expected
to manifest itself when the theory was pushed to describe the actual collision of black holes.
The subsequent discovery that the radiation from the merger evolved very simply, smoothly connecting the
amplitude and phase of the inspiral to those of the ringdown across all of the relevant parameter space,
was a validation of two complementary efforts predicated on the assumed smallness of nonlinear effects
throughout the entire coalescence -- the close-limit approximation \cite{Price:1994pm} culminating in the Lazarus
project \cite{Baker:2001sf}, and the Effective One-Body (EOB) approach \cite{Buonanno:1998gg,Buonanno:2000ef}. However, although the 
smoothness of the merger has made it possible to create analytical models by 
extending post-Newtonian results to include free parameters, and tuning those to NR results
[as is done in both EOB and the inspiral-merger-ringdown
phenomenological (IMRPhenom) family of models \cite{Ajith:2007qp}],
there is currently no accurate model of the merger that is constructed analytically from first principles, rather than through a fit to NR.

The phenomenological approach to modeling mergers
has achieved great success in estimating the parameters of the black-hole binaries (BHBs) observed by the Advanced 
Laser Interferometer Gravitational-Wave Observatory (LIGO) \cite{LIGODA}.
However, the LIGO-Virgo Collaboration found
that a non-negligible subset of parameter space would be limited by systematic modeling errors even at current sensitivities \cite{systematics}.
Future upgrades to Advanced LIGO,
as well as space-based instruments like LISA, will detect signals with substantially larger signal-to-noise
ratios \cite{McWilliams:2010eq}, placing far more stringent requirements on the systematic modeling errors that can be tolerated. For more subtle
measurements, such as tests of general relativity, the most useful observations by far will be the loudest; events
with signal-to-noise ratios in the thousands will require modeling errors hundreds of times smaller than what has been required 
to date. Such requirements may be beyond the current capabilities of NR, let alone phenomenological models tuned 
to NR results.
 
We emphasize that, just as has been the case for all BHB detections to date, the late inspiral
and merger-ringdown
is expected to constitute the majority of the signal-to-noise ratio for most expected sources for both ground- and space-based
observatories \cite{McWilliams:2010eq}. For concreteness, we refer to the ``merger-ringdown'' as the part of the waveform occurring 
at and after the
time of peak amplitude for the strain $h$, noting that the time of peak strain, the time of peak amplitude for the Weyl scalar $\psi_4$ 
(the primary output of most NR codes, which is proportional to $\ddot{h}$),
as well as
the time of formation of a common apparent horizon in NR simulations, all occur within a few $M$
of each other 
(i.~e.,~ of order the light crossing time of the final black hole), where we will use geometrized units
where $G=c=1$ throughout.
We refer to the ``late inspiral''
as the part of the waveform sourced by the system after it reaches the innermost stable circular orbit (ISCO) of
the final merged black-hole spacetime, but before it reaches the light ring.

We will show that the spacetime of the
final merged remnant provides the most useful equivalent one-body system for describing the post-ISCO dynamics.
Since the background spacetime on which we find a perturbative solution is the state that the system is known to approach 
at later times, we refer to this approach as the Backwards One-Body (BOB) method. 
The BOB method does not include any
phenomenological degrees of freedom, yet it performs as well as the most accurate models that have been
tuned to NR results; in fact, as we will show, the BOB method agrees with NR results to within those results' own stated
uncertainties throughout the entire merger-ringdown, and maintains accuracy back to the ISCO
of the equivalent single black-hole system over a large portion of parameter space.
The high degree of fidelity of this model strongly suggests that the description
of the binary that motivates the model is providing a physically meaningful description of the late stage dynamics of merging
BHBs.

\paragraph{Physical description of mergers.---}
\label{sec:physics}
It has previously been noted \cite{Goebel} that, within the eikonal
approximation where $\ell\, \geq\, |m| \gg 1$, with $\ell$ and $m$ being the harmonic indices, 
the gravitational-wave emission of a single perturbed black hole is well described
by the properties of null geodesics on unstable circular orbits at the black hole's light ring.
These quasinormal
modes (QNMs) should describe the end state of a BHB merger, so that the emission at late times must
in some way relate to the dynamics of null rays at the light ring. However, it has also been
argued that the peak in the gravitational-wave amplitude corresponds, in the EOB description, to the 
perturber crossing the light ring of the effective single black-hole spacetime \cite{Buonanno:2000ef}, so that the emission
at the moment of merger should also correspond in some way to the dynamics of null rays at the light ring. This begs
an obvious question: How can the merger waveform, as well as the waveform at a time well after the merger, {\bf both} correspond 
to disturbances at the light ring?

To understand this dichotomy, we interpret the sequence of events in the following way. First, we consider an effective
single black-hole spacetime with an inspiraling perturber.
As the perturber approaches the light ring of the black hole,
most of the gravitational-wave emission that will reach a distant observer is actually being reflected by the curvature
potential of the black hole, rather than arriving directly from the perturber \cite{Buonanno:2000ef,Damour:2007xr,Racine:2008kj,Nichols:2010qi,Barausse:2011kb}. This emission will occur at harmonics
of the perturber's instantaneous orbital frequency and will spiral outward 
along the outgoing geodesic path for escaping null particles with the same angular momentum as the perturber.
Because radiation reaction has a negligible effect on the dynamics inside ISCO \cite{Buonanno:2000ef,Ori:2000zn}, the point particle
follows a timelike geodesic path. As the perturber passes beyond the light ring, 
most of the radiation that it sources directly falls into the black hole; however,
a range of spacetime disturbances with higher frequencies
is also generated at the light ring, either by the passage of the perturber or through a nonlinear response to the emission at lower
frequencies. These higher frequencies span from the
perturber's frequency up to the null circular orbital frequency at the light ring, with gravitational-wave emission
being sourced at multiples of these frequencies.
Higher-frequency null rays spend more time orbiting the system, in addition to any potential intrinsic delay
in generating higher-frequency emission,
so that higher-frequency gravitational waves will reach distant observers at later times, in direct analogy to the behavior
of light escaping a collapsing star \cite{Ames}.
The frequency of orbiting perturbations asymptotes to the null circular orbit frequency, since those perturbations 
orbit the black hole indefinitely. 
In Supplemental Material \cite{supp}, we include an illustration to further clarify our description of the merger dynamics.

\paragraph{Merger amplitude.---}
\label{sec:amp}
The frequencies of the QNMs of a single perturbed black
hole closely match the corresponding harmonics of the orbital frequency for a null geodesic circling the
light ring, and the decay rate of the amplitude corresponds to the Lyapunov coefficient
characterizing the rate of divergence of nearby null geodesics \cite{Ferrari}.
This correspondence is well motivated in the geometric optics limit where $\ell\, \geq\, m \gg 1$ but provides
accurate predictions even for small $\ell$ and $m$.
The QNM family of exponentially decaying sinusoids can therefore
be found by calculating the behavior at late times of a bundle of null geodesics, known as a null congruence, that has diverged from the light ring \cite{Yang}.
However, if we trace the behavior of the congruence back to the point where the bundle converges,
which one would expect to be associated with the peak waveform amplitude, 
then we can predict the behavior of the amplitude at earlier times.

To accomplish this, we follow a similar approach to Ref.~\cite{Ferrari},
in that we consider a set of
geodesics perturbed away from light-ring orbits, except that we consider perturbations in all directions, whereas
past authors have focused on perturbations within the equatorial plane. In other words, for
geodesics described by the set of coordinates \{$t$, $r$, $\theta$, $\phi$\}, we express their evolution at leading order by
\bea
t &=& \tpeak + \eta+\epsilon \mathfrak{h}(t-\tpeak)\,, \nonumber \\
r &=& \rlr[1+\epsilon \mathfrak{f}(t-\tpeak)]\,, \nonumber \\
\theta &=& \frac{\pi}{2}[1+\epsilon \mathfrak{p}(t-\tpeak)]\,, \nonumber \\
\phi &=& \omega[t+\epsilon \mathfrak{g}(t-\tpeak)]\,, 
\label{eq:pert}
\eea
where $\tpeak$ is the time when the congruence converges, corresponding to the peak waveform amplitude, $\eta$ is an affine parameter, $\epsilon$ is a small dimensionless order-counting parameter, $\rlr$ is the light-ring radius,
$\omega$ is the orbital
frequency of the geodesic, and $\mathfrak{f}$, $\mathfrak{g}$, $\mathfrak{h}$, and $\mathfrak{p}$ are
functions determined from the requirements that the perturbed orbits are still null geodesics, and that
$\mathfrak{f}(0)=\mathfrak{g}(0)=\mathfrak{h}(0)=\mathfrak{p}(0)=0$. We note that in Ref.~\cite{Ferrari}, 
$\theta$ is held fixed at $\pi/2$ while the other coordinates are
perturbed. This difference is minor when considering QNMs, and amounts to a different
convention for the Lyapunov coefficient, but when considering the evolution of the amplitude at times
as early as the peak, this difference is more significant. The resulting perturbation functions are given by
\bea
\mathfrak{f} &=& \sinh[\gamma (t-\tpeak)]\,, \nonumber \\
\mathfrak{g} &=& 0\,, \nonumber \\
\mathfrak{h} &=& 2\frac{\omega}{\gamma^2}\sqrt{\frac{3M}{\rlr}}\{1-\cosh[\gamma (t-\tpeak)]\}\,, \nonumber \\
\mathfrak{p} &=& 0\,,
\label{eq:func}
\eea
where $\gamma$ is the Lyapunov exponent of the congruence, and corresponds in the wave picture to the inverse
damping time of the amplitude.
In particular, note that to leading order in $\epsilon$, we find perturbed geodesics do not evolve in the $\theta$
direction.  This result might at first appear to validate fixing $\theta=\pi/2$ as in Ref.~\cite{Ferrari}, since we arrive
at the same result for the differential cross-sectional area of
the congruence, namely, that
\beq
d\mathcal{A} = d\mathcal{A}_0 \cosh[\gamma (t-\tpeak)] = \pi \rlr dr d\theta\,.
\eeq
However, this result is potentially misleading, as the expansion occurs only along the radial direction, and not in the polar direction,
so $d\theta$ is constant, and only $dr \propto \cosh[\gamma (t-\tpeak)]$ evolves with time.
Since $dr/d\eta = dr/dt + \mathcal{O}(\epsilon^2)$, there is no need to distinguish between time and the affine parameter at leading order,
and we need only to focus on the behavior of $r$ in Eq.~\eqref{eq:pert} to determine the behavior of the waveform amplitude.

The dimensionality of the expanding null congruence,
and its relationship to the wave amplitude within the geometric optics approximation, is therefore modified relative
to the case of expansion in empty space. In particular, the transport equation
relating the cross-sectional area and the waveform amplitude $A$ becomes
\bea
k^{\mu}\partial_{\mu}(dr A)&=&0\,, \nonumber \\
\therefore A&=&\Apeak\sech[\gamma (t-\tpeak)]\,.
\label{eq:amp}
\eea
We note that this conclusion differs from previous treatments of QNMs
in the geometric optics limit (e.~g.,~Ref.~\cite{Yang}) that applied the result for expansion in empty space, $k^{\mu}\partial_{\mu}(d\mathcal{A}^{1/2} A)=0$ \cite{Misner73}, which is not valid for these orbits and would lead one to
conclude in our case that $A=\Apeak\sech^{1/2}[\gamma (t-\tpeak)]$ rather than the correct result given in Eq.~\eqref{eq:amp}.
This discrepancy highlights an important-but-subtle distinction, that although the actual radiation in the far field should of course 
propagate as if it were in empty space, the duality between QNMs and null light-ring orbits is a feature of the Kerr solution
in the near zone (i.~e.,~it is a feature of the \emph{source}, not the emission), 
and therefore the correct strong-field effects on wave propagation should be taken into account.  

The amplitude $A$ could, in principle, describe any derivative or integral of the gravitational-wave strain. However, given our goal
of developing a model that can be extended to times before the peak (and ideally back to the ISCO),
we next considered which derivative of strain would have an amplitude best described by Eq.~\eqref{eq:amp}
at $t<\tpeak$. 
We will present the full details of the calculation in followup work, but in summary, 
we solved an approximation to the sourceless Zerilli equation \cite{Zerilli:1970se} that describes
the scattering of gravitational perturbations by a black hole to first order in the black-hole spin.
Previous work has shown that, just prior to the merger, the dominant contribution to the gravitational-wave emission comes
from gravitational perturbations scattering off of the curvature potential rather than arriving directly from the
effective perturber \cite{Buonanno:2000ef,Damour:2007xr,Racine:2008kj,Nichols:2010qi,Barausse:2011kb},
so that the Zerilli equation can be used to describe the emission
during this time.
We replaced the exact curvature potential that appears in Ref.~\cite{Zerilli:1970se} with a negated Poschl-Teller potential \cite{Poschl},
an approximation that has been used successfully to find analytical solutions
for the QNM frequencies for nonspinning and slowly spinning systems \cite{Ferrari},
and found that
\beq
|\psi_4| = \Apeak \sech[\gamma (t-\tpeak)]
\label{eq:psi4amp}
\eeq
satisfies Eq.~\eqref{eq:amp} for $t<\tpeak$ to $\mathcal{O}[(t-\tpeak)^4]$, better than any other strain derivative.
Since $|h|\approx |\psi_4|/\omega^2$ for quasicircular systems, we 
can also combine Eq.~\eqref{eq:psi4amp} with an analytical model for $\omega$ to define an analytic model for the strain amplitude. 

We note that, although our results from this section formally hold only for $\ell=|m|$ modes, a numerical
study of geodesic deviation for nonequatorial (i.~e.,~$\ell > |m|$) modes suggests that radial expansion
dominates polar expansion in all cases. Nonetheless,
the NR data are generally of poor quality for these modes and are known to suffer from mode mixing \cite{Kelly},
so in Supplemental Material \cite{supp}, we show the agreement of the model with the loudest nonmixed mode with $\ell > |m|$ (specifically, $\ell=2$, $|m|=1$) but leave
a more detailed study for future investigation.

\paragraph{Phase evolution.---}
\label{sec:phase}
With a model for the amplitude of $\psi_4$ in hand, we can now turn to modeling the phase of $\psi_4$, and
recovering the strain from these quantities.  To this end, we follow a similar approach to that employed by the author
and collaborators in Ref.~\cite{Baker:2008mj}, where a phenomenological model for the frequency was developed,
and a relationship between amplitude and frequency was derived to complete the model.  We instead have
developed a first-principles model for the amplitude, but we can apply the same relationship as in Ref.~\cite{Baker:2008mj}
to calculate the frequency (and subsequently the phase) from the amplitude.

Specifically, we can relate the amplitude and frequency of the news, $\mathcal{N}_{\ell m}= \dot{h}_{\ell m}$, using
\beq
|\mathcal{N}_{\ell m}|^2 = 16\pi \xi_{\ell m} \Omega_{\ell m} \dot{\Omega}_{\ell m} = 8\pi \xi_{\ell m} \frac{d}{dt}(\Omega_{\ell m}^2)\,,
\label{eq:news}
\eeq
where $\xi_{\ell m} \equiv m^2 \frac{dJ_{\ell m}}{d\Omega_{\ell m}}$ 
was shown in Ref.~\cite{Baker:2008mj} to remain constant throughout the late inspiral and merger-ringdown, and indeed would be
expected to trend to a constant due to the exponential asymptotic approach of both $J_{\ell m}$ and $\Omega_{\ell m}$ to their final
constant values, with the $e$-folding timescale of both set by the final black-hole damping time. 
$\Omega_{\ell m}$ represents the orbital frequency, first of the perturber, and subsequently
of the inferred spacetime perturbations orbiting near the light ring and continuing to source gravitational-wave 
emission. We note that the different $\Omega_{\ell m}$'s should be equal when sourced by a single 
perturber, but can differ from each other once the emission decouples from a single source. We will drop the
subscripts in what follows for notational simplicity and lack of ambiguity, but
before doing so, we emphasize that at no point do we enforce equality of the different $\Omega_{\ell m}$'s.
Indeed, we find that the different $\Omega_{\ell m}$ curves for the $\ell=|m|$ modes are quite similar and
their amplitudes peak at nearly identical times,
whereas for the $\ell > |m|$ modes, $\Omega_{\ell m}$ begins to notably differ through the merger, and the amplitude
peaks at different times, consistent with previous studies \cite{Baker:2008mj}.

Since $|\psi_4|^2 = \frac{d}{dt}|\mathcal{N}|^2 + m^2 \Omega^2
|\mathcal{N}|^2 \approx m^2 \Omega^2 |\mathcal{N}|^2$ due again to quasicircularity, 
we can insert Eq.~\eqref{eq:psi4amp} into
Eq.~\eqref{eq:news}, separate the $\Omega$ and $t$ variables, and integrate to find
\beq
\Omega = \left\{\Omega_0^4 + k \left[\tanh\left(\frac{t-\tpeak}{\tau}\right)-\tanh\left(\frac{t_0-\tpeak}{\tau}\right)\right]\right\}^{1/4}\,,
\label{eq:om}
\eeq
where the constant $k$ is given by
\beq
k=\left(\frac{\Omega_{\rm QNM}^4-\Omega_0^4}{1-\tanh\left[(t_0-t_p)/\tau \right]}\right)\,,
\eeq
where $\tau = \gamma^{-1}$ is
the damping time, and $\Omega_{\rm QNM} = \omega_{\rm QNM}/m$ is the inferred asymptotic orbital frequency of light-ring perturbations
sourcing QNMs with frequency $\omega_{\rm QNM}$.
We note that only relative time shifts are physically meaningful, so that either $\tpeak$ or $t_0$ can be freely chosen. The parameters 
$\tpeak-t_0$, $\Apeak$, and $\Omega_0$ can be set by enforcing continuity with the inspiral of the amplitude, frequency, and either
of their derivatives. 
Since phase agreement is generally more important, we opt to enforce continuity in $\dot{\Omega}=\frac{k}{4\tau \Omega^3}\sech^2\left(\frac{t-\tpeak}{\tau}\right)$,
so that
\beq
t_0=\tpeak-\frac{\tau}{2} \ln \left(\frac{\Omega_{\rm QNM}^4-\Omega_0^4}{2\tau \Omega_0^3 \dot{\Omega}_0}-1\right)\,,
\eeq
where $\Omega_0$ and $\dot{\Omega}_0$ are the orbital frequency and frequency derivative, respectively, at the
transition from inspiral. We can additionally relate the amplitude at the transition point to the amplitude
at the peak using $A_0\equiv \Apeak \sech[(t_0-\tpeak)/\tau]$. We note 
that we could alternatively eliminate $t_0$ as a parameter altogether in favor of fixing $\Omega(t\rightarrow -\infty)$ to
a particular value, with 0 or $\Omega_{\rm ISCO}$ being physically motivated choices; we will explore these possibilities further in future work.  

Finally, we integrate Eq.~\ref{eq:om} to find the phase,
\bea
\Phi = \displaystyle\int_0^t \Omega\, dt'&=& \arctan_+ + \arctanh_+ - \arctan_- - \arctanh_- \,, \nonumber \\
\atanhopt_{\pm} &\equiv& \kappa_{\pm} \tau \left[\atanhopt\left(\frac{\Omega}{\kappa_{\pm}}\right)
- \atanhopt\left(\frac{\Omega_0}{\kappa_{\pm}}\right)\right] \,, \nonumber \\
\kappa_{\pm} &\equiv& \left\{\Omega_0^4 \pm k \left[1 \mp \tanh\left(\frac{\tstart-\tpeak}{\tau}\right)\right]\right\}^{1/4}\,.
\label{eq:phi}
\eea
Since Eqs.~\eqref{eq:om} and \eqref{eq:phi} represent the rotation of the source, the frequency and phase are simply given as
$\omega_{\ell m} = m\Omega$ and $\phi_{\ell m} = m\Phi$, respectively. 

\paragraph{Results.---}
\label{sec:result}
To complete our first-principles model based on the final state of the system, we require a method for predicting the
final mass and spin of the merger remnant based on the initial conditions of the system.  A considerable amount of work
has been done on generating fitting formulas to suites of NR simulations, so those fits could be
used for this purpose.  However, in the interest of generating a waveform that does not rely on NR
results in any way, we can instead apply the first-principles approach used in Ref.~\cite{Buonanno:2007sv}
but supplemented to include the change in mass due to the loss of binding energy through gravitational radiation \cite{Bardeen}.
We include full details of this model and a comparison to a numerical-relativity-based fit \cite{Kellyb} in Supplemental Material \cite{supp}.

\begin{figure}[ht!]
\includegraphics[trim = 5mm 0mm 0mm 0mm, clip, width=.48\textwidth, angle=0]{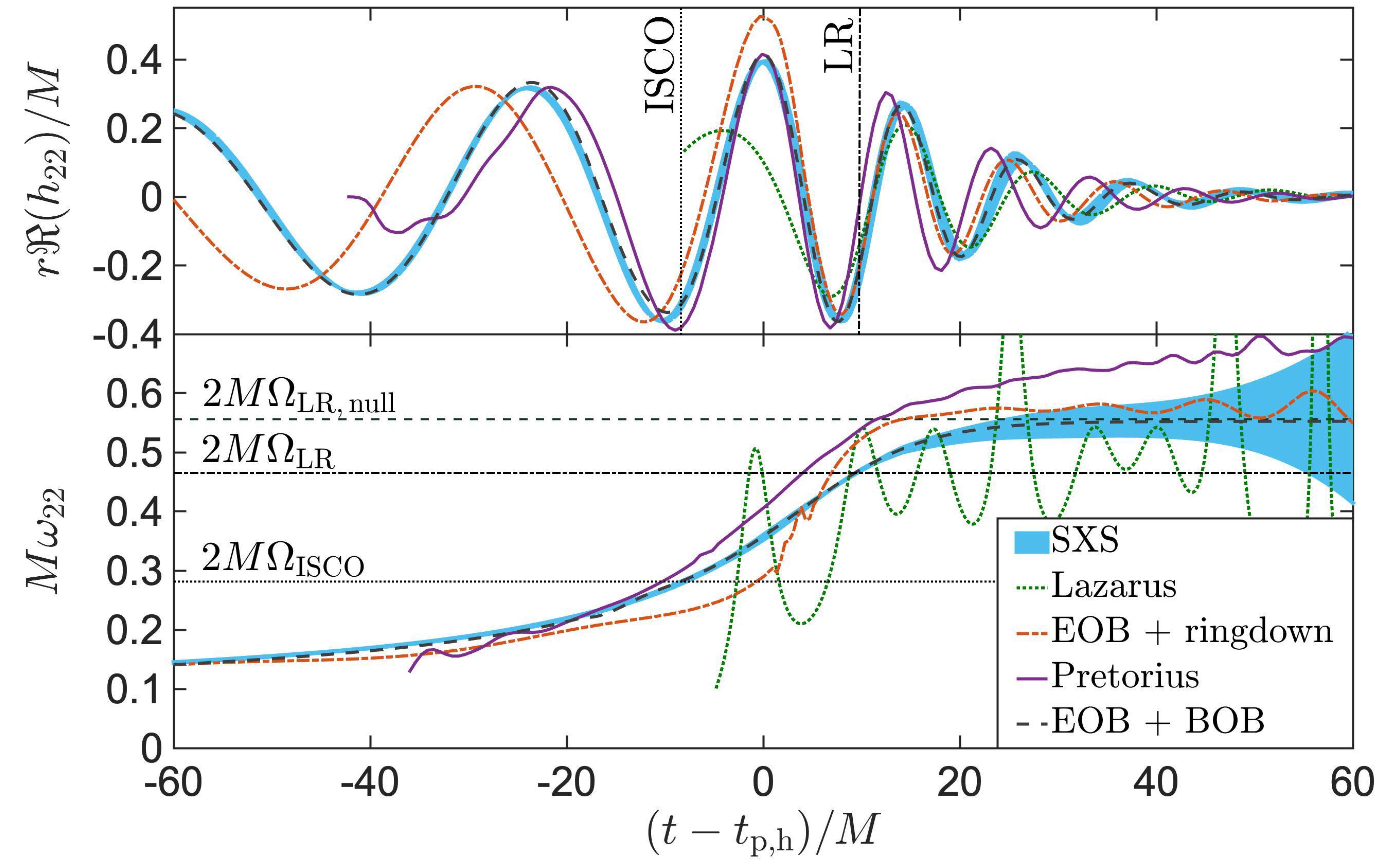}
\caption
{
Historical comparison of waveform predictions for the dominant $\ell=m=2$ mode of strain for an equal mass, nonspinning merger, including
the Lazarus project \cite{Baker:2001sf} [(green) dotted line], uncalibrated EOB attached to a ringdown \cite{Buonanno:2006ui} [(red) dash-dotted line],
the first stable evolution of a binary merger in numerical relativity
by Pretorius \cite{Pretorius:2005gq} [(magenta) solid line], and simulations by the SXS collaboration \cite{Mroue} [(cyan) shaded region].
We note that the times and extraction distances for all the waveforms are rescaled by their own estimates of the final mass and
that the waveforms from Refs.~\cite{Pretorius:2005gq,Buonanno:2006ui} have small nonzero initial spins. We estimated an uncertainty interval for the SXS waveform
(SXS:BBH0001 from the online waveform catalog \cite{catalog}) 
by combining (in quadrature) the
numerical error derived from the multiple
available resolutions, the extrapolation error derived from the many available extraction radii, and
systematics from residual eccentricity as estimated from a second available SXS simulation (SXS:BBH0002) that used different initial
data. When we transition from the same uncalibrated EOB inspiral to the BOB model $5M$ after it reaches
twice the ISCO orbital frequency of the merged remnant,
so that we are replacing most of the EOB extrapolation to the light ring and subsequent ringdown attachment with BOB
[(black) dashed line], the resulting waveform agrees with SXS to within
SXS's uncertainties throughout the merger ringdown and backwards in time beyond the ISCO.
For reference, we also show the times (top) and frequencies (bottom) corresponding to the SXS waveform crossing twice the ISCO frequency and twice the light-ring (LR) frequency of the infalling perturber, as well as the frequency (bottom) for crossing
twice the circular light-ring frequency (LR, null). All waveforms are aligned at the time of peak strain amplitude, $t_{p,h}$.
}
\label{fig:q1comp}
\end{figure}

We show the result of this approach in Fig.~\ref{fig:q1comp}, where we attached our fully NR-independent BOB model to an EOB inspiral
that follows the methodology referenced in Ref.~\cite{Buonanno:2006ui} and is not calibrated to NR results. We aligned the data at the time of
peak strain amplitude, $t_{p,h}$, so we note that $t_{p,h} \approx t_p + 10 M$ for this case. The attachment was done by 
first enforcing agreement with the EOB amplitude, frequency, and frequency derivative at
a time $5M$ after it reaches the ISCO frequency. In principle, this alignment could be done anywhere between ISCO and the
light ring; however, since we want to minimize our reliance on EOB inside the ISCO, but our physical model suggests that
$\Omega$ should asymptote to a value comparable to the ISCO value at early times, we choose a time just after the ISCO crossing to enforce continuity. We then smoothly
transition between the two models using a raised cosine over a time window of width $5M$ starting at $t-\tpeak=-20M$, which was chosen
so that the transition to BOB would begin when the EOB inspiral
reaches the ISCO frequency of the merged remnant. We compare this result
to various past waveform predictions, and show that BOB is not only a dramatic improvement over historic alternatives to full NR, but it
actually agrees with the state of the art in NR, as represented by the latest Simulating eXtreme Spacetimes (SXS) result, to within our estimate of SXS's own uncertainties. In Supplemental Material \cite{supp}, we show additional comparisons between the BOB model and a set
of NR waveforms that span the full range of physical parameter space available from the SXS catalog of waveforms \cite{catalog},
which show that the outstanding agreement between BOB and NR extends across the full parameter space.

\vspace{0.2in}
\begin{acknowledgments}
This work is partially supported by the National Science Foundation under Grant No. OIA-1458952. 
We thank Alessandra Buonanno and Lionel London for comments on the manuscript and 
Neil Cornish for suggesting the name ``BOB'' over drinks in Albuquerque.
\end{acknowledgments}

\section*{Supplemental Material}

\begin{figure*}[ht!]
\begin{center}
\includegraphics*[trim = 0mm 0mm 0mm 0mm, clip, width=.50\textwidth, angle=0]{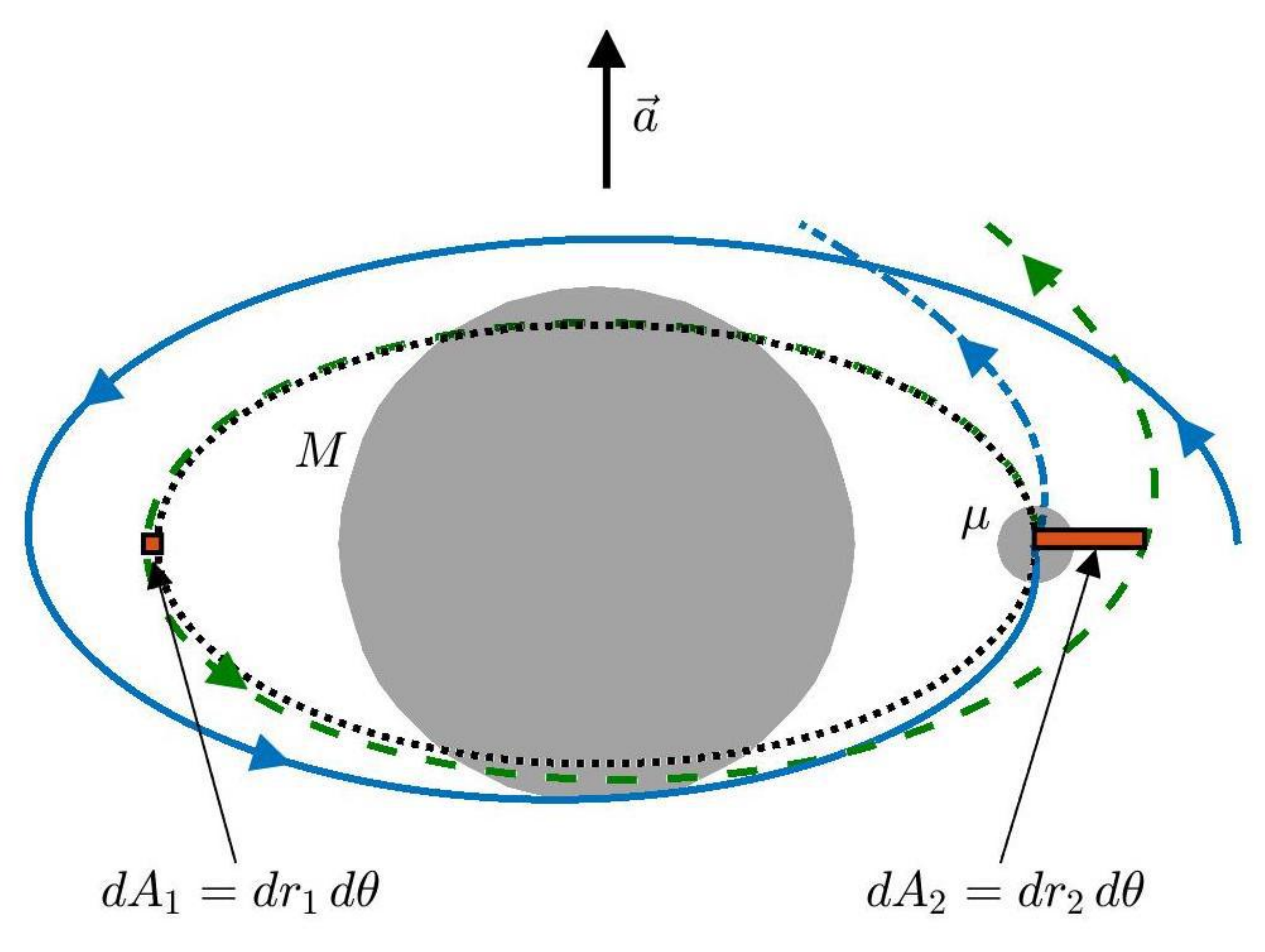}
\caption
{
The behavior of null rays sourced by a perturber with mass $\mu$ near the light ring
of a black hole with mass $M$ and spin $a$
corresponding to the final mass and spin of a merged system.
The perturber
follows a timelike
geodesic path ((blue) solid line) from the ISCO through the light ring, where
it sources gravitational radiation at harmonics of its instantaneous orbital frequency. The radiation
travels along
the outgoing geodesic path for escaping null particles with the same angular momentum as the perturber
((blue) dash-dotted line).  A
range of spacetime disturbances with higher frequencies
are also generated by the passage through the light ring ((green) dashed line), spanning from the
perturber's frequency up to the circular null frequency at the light ring ((black) dotted line).
Null geodesics that are perturbed from the light ring diverge in the radial direction, but not in the polar direction,
as illustrated by the (red) boxes showing the area of a null bundle, $dA_1$, and its area at a later time, $dA_2$.
}
\label{fig:kerrgeo}
\end{center}
\end{figure*}

In Fig.~\ref{fig:kerrgeo}, we illustrate our physical description of the merger dynamics as being equivalent to an effective
single-black hole spacetime with an inspiraling perturber. This approach is similar in spirit to the EOB method, but with
the key difference that our effective black hole corresponds to the final merged remnant, rather than corresponding
to an effective (post-)Newtonian system that most accurately describes the
state of the system at early times, as is the case in EOB. The resulting gravitational-wave emission 
is most easily understood through the geometric optics-motivated
description of rays of gravitational waves. The perturber sources a bundle of emission,
with the density of rays decreasing as we move
away from the center of the bundle. In addition to the rays being less dense (which
corresponds to a decreasing amplitude in the wave picture),
the rays themselves have trajectories with ever smaller ratios of radial to azimuthal momentum, hugging
progressively closer to the light ring, being bent further and further around the black hole, and consequently
taking longer to reach a distant observer. The later null rays have essentially all of their momentum in
the azimuthal direction, and are therefore orbiting very near the circular null orbit, whose frequency
is known to correspond to that of quasi-normal mode emission.

To formulate a truly first-principles model based on the final state of the system requires a method for predicting the
final mass and spin of the merger remnant based on the initial conditions of the system.  A considerable amount of work
has been done on generating fitting formulae to suites of numerical relativity simulations, so those fits could be
used for this purpose.  However, in the interest of generating a waveform that does not rely on numerical relativity
results in any way, we have employed a first-principles approach to predicting the end state.  In \cite{Buonanno:2007sv}, the authors
developed a model for the final spin of a merged remnant that performs fairly well without relying on numerical relativity results, by
taking the individual black hole spins, whose magnitudes remain constant throughout the inspiral to excellent
approximation, and adding them to the orbital angular momentum of a test particle orbiting the final merged black hole
at its ISCO:
\bea
\hat{a}_f = \frac{L_{\rm orb}(r_{\rm ISCO},\,\hat{a}_f)}{M_f^2} &+& \frac{\hat{a}_1}{4}\left(1+\sqrt{1-4\eta}\right)^2 \nonumber \\
&+& \frac{\hat{a}_2}{4}\left(1+\sqrt{1-4\eta}\right)^2\,,
\eea
where $\hat{a}_f \equiv a_f/M_f \equiv S_f/M_f^2$ is the final black hole's dimensionless Kerr parameter, $\hat{a}_i \equiv a_i/m_i \equiv S_i/m_i^2$ are the Kerr parameters of the individual inspiralling black holes, and $\eta \equiv m_1m_2/M_f^2$ is the
symmetric mass ratio.  

The authors of \cite{Buonanno:2007sv} acknowledge neglecting
the change in system mass due to gravitational radiation.  However, this change in mass could (and, as we know from numerical
relativity, does)
change the final spin at the several percent level, exceeding the uncertainties achieved by numerical
relativity-based fits. However, this change can be included to the same level of approximation as the rest of the expression.
Specifically, if we consider an effective system wherein a test mass $\mu$ inspirals toward a black hole of mass $M$, such that the total
ADM mass $M_{\rm ADM}=M+\mu = M(1+\eta)$, then the mass remaining bound in the system when the test mass reaches ISCO is
\beq
M_f \equiv M+\mu+E_{\rm bind} = M+E_{\mu,\,{\rm tot}} = M+\hat{E} \mu = M(1+\hat{E} \eta)\,,
\eeq
where $E_{\rm bind}$ is the (negative-valued) gravitational binding energy at ISCO, $E_{\mu,\,{\rm tot}}$ is the total effective mass of the perturber (including rest mass and binding energy), and the fraction of the test mass radiated away by the time it reaches the ISCO is given by \cite{Bardeen} 
\beq
\hat{E} \equiv \frac{1}{r_{\rm ISCO}^{3/4}}\left(\frac{r_{\rm ISCO}^{3/2} - 2M_f r_{\rm ISCO}^{1/2} + M_f^{1/2}a_f}{\sqrt{r_{\rm ISCO}^{3/2} - 3M_f r_{\rm ISCO}^{1/2} + 2M_f^{1/2}a_f}}\right)\,.
\eeq
Therefore, we arrive at the final result that the final mass is given as a fraction of the ADM mass by
\beq
\frac{M_f}{M_{\rm ADM}} = \frac{1+\hat{E} \eta}{1+\eta}\,,
\eeq
and that, with the conventional choice that $M_{\rm ADM}\equiv 1$, this expression for the final mass can then be used in the calculation of the final spin.
 
\begin{figure*}[ht!]
\begin{center}
\includegraphics*[trim = 0mm 0mm 0mm 0mm, clip, width=.5\textwidth, angle=0]{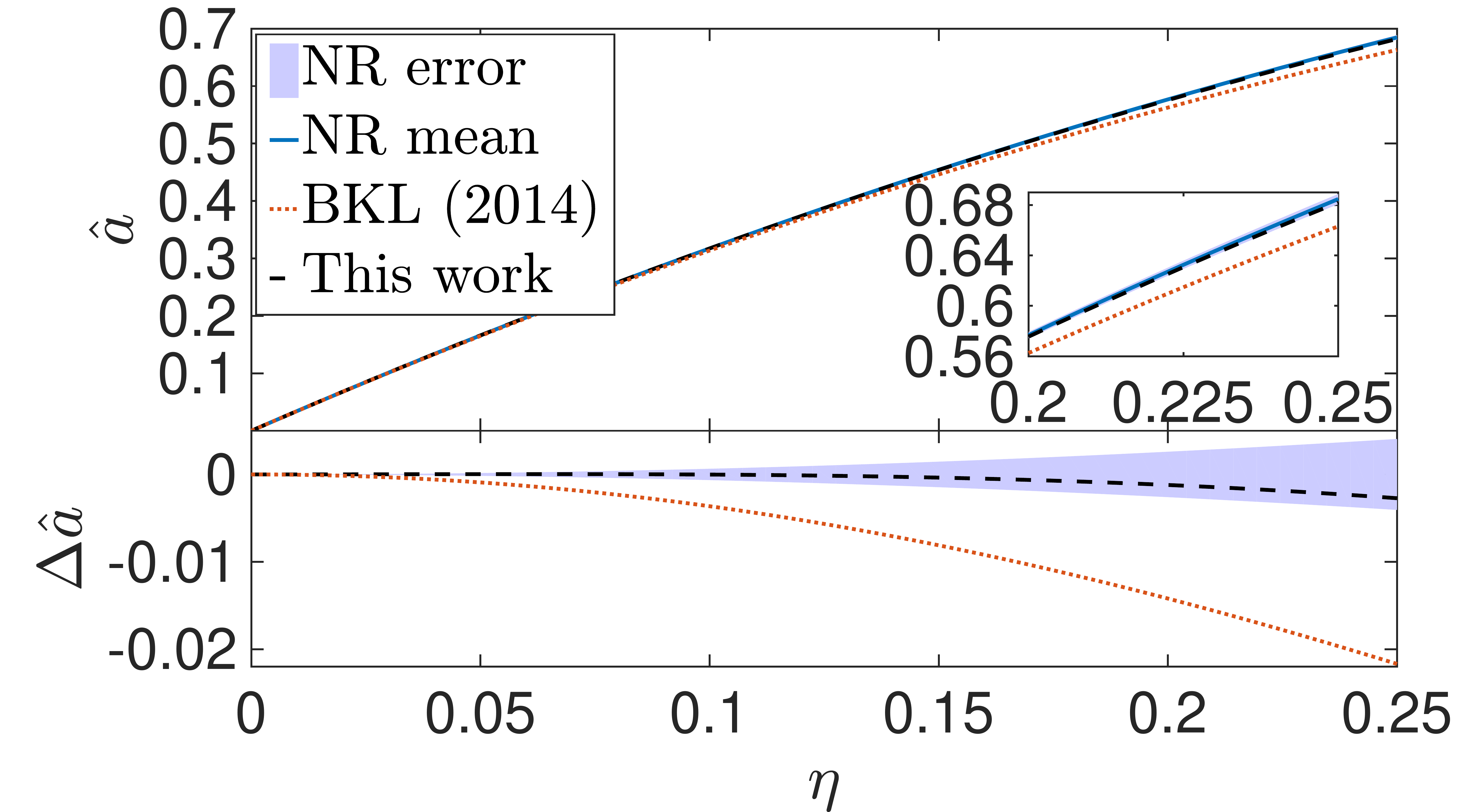}
\caption
{
Comparison between our first-principles prediction of the spin of the merged black-hole remnant ((black) dashed lines), and a fit to numerical relativity results from \cite{Kellyb} ((blue) solid line), including their estimated uncertainty ((light blue) shaded region). We also show the prediction from \cite{Buonanno:2007sv} ((red) dotted line), which neglects the mass loss due to gravitational radiation.
}
\label{fig:ahat}
\end{center}
\end{figure*}

We note that this estimate of the final mass should, if self-consistent, only include the radiative losses up until ISCO, and should therefore
not include radiative losses during merger, which are known to approximately scale with $\eta^2$, but can only be fully determined by numerical
relativity. Nonetheless, since we must only include losses up until ISCO, we can avoid any appeal to numerical relativity results.
In Fig.~\ref{fig:ahat}, we compare the resulting prediction for the final black-hole spin for the full range of nonspinning binaries, noting that
our proposed modification to the model in \cite{Buonanno:2007sv} affects only the contribution of orbital angular momentum to the final spin, and so
is fully tested by comparison to nonspinning numerical relativity results.

\begin{figure*}[ht!]
\begin{center}
\includegraphics*[trim = 5mm 0mm 0mm 0mm, clip, width=\textwidth, angle=0]{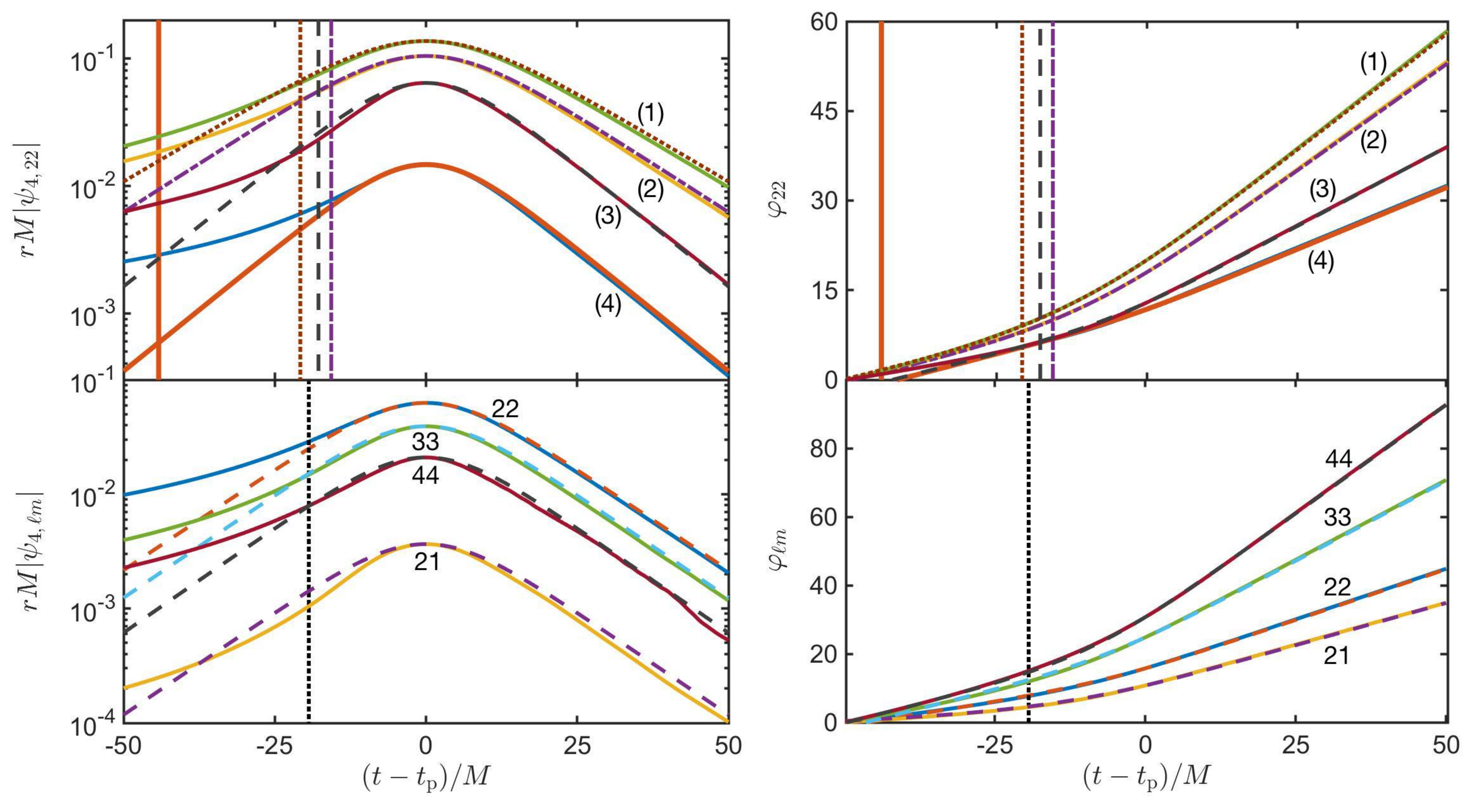}
\caption
{
Comparison of $\psi_4$ amplitudes (left panels) and phases (right panels) between BOB (solid lines) and SXS (various) for several cases.
We align the phases with each other at the peak time, then shift both phases together so that the SXS phases
are zero at $t-\tpeak=-50\,M$. We normalize the BOB amplitudes to agree with the corresponding SXS data at the peak,
although we note that we could also determine the normalizations from continuity with a generic spinning 
inspiral model, in analogy to our approach to the equal mass nonspinning case in Fig. 1.
In the top panels, we compare the dominant $\ell=m=2$ modes for (1) an equal mass binary with aligned spins of 0.9
on each black hole (SXS:BBH0160), (2) a precessing 3:2 mass ratio binary, the larger mass having an aligned spin of 0.991 and the smaller
having a randomly misaligned spin with total magnitude 0.2 (SXS:BBH0179), (3) a precessing equal mass binary with identically misaligned
spins each of magnitude 0.6 (SXS:BBH0161), and (4) a 10:1 mass ratio nonspinning binary (SXS:BBH0303). For each case, we also show the ISCO
time for BOB, using the corresponding line style (and color). In the bottom panel, we show the $\ell=2$, $m=1$ (21),
$\ell=2$, $m=2$ (22), $\ell=3$, $m=3$ (33), and $\ell=4$, $m=4$ (44) harmonics for a 3:1 mass ratio binary with
each black hole having anti-aligned spins of $-0.5$ (SXS:BBH0046), with the ISCO time for BOB shown for reference.
The phase difference between the BOB model and the SXS simulations is $\mathcal{O}$(0.1 rad) for all times $t-\tpeak \gsim -20 M$ in all cases,
which is comparable to the phase uncertainties of the simulations.
The amplitudes are likewise in agreement to within the SXS uncertainties over the same time interval.
}
\label{fig:psicomp}
\end{center}
\end{figure*}

In Fig.~\ref{fig:psicomp}, we compare the amplitudes and phases between BOB and various SXS example cases. For this comparison, we
used the final masses and spins quoted in the SXS catalog for each case, and compared $\psi_4$ instead of strain,
so that the comparison would focus only on intrinsic differences
in the models. We chose cases that represented extremes in mass ratio, spin magnitude, and spin precession available within the SXS catalog,
finding that BOB agreed with SXS to within SXS's uncertainties in both phase and amplitude for all times $t-\tpeak>-20\,M$. This interval
covered times as early as the merger-remnant ISCO for all but one case that we considered, so that BOB could in principle be combined with
uncalibrated EOB to form a complete and accurate model for those cases. The one exception, a 10:1 mass ratio system, represents a region
of parameter space where further improvements to BOB and/or calibration of EOB to NR is still required to reach the desired accuracy.

\begin{thebibliography}{10}
  
\bibitem{Pretorius:2005gq}
F. Pretorius, Phys. Rev. Lett. {\bf 95},  121101  (2005).

\bibitem{Campanelli:2005dd}
M. Campanelli, C.~O. Lousto, P. Marronetti, and Y. Zlochower, Phys. Rev. Lett.
  {\bf 96},  111101  (2006).

\bibitem{Baker:2005vv}
J.~G. Baker {\it et~al.}, Phys. Rev. Lett. {\bf 96},  111102  (2006).

\bibitem{NRreview}
L. Lehner and F. Pretorius, Annu. Rev. Astron. Astrophys. {\bf 52},  661
  (2014).

\bibitem{Price:1994pm}
R.~H. Price and J.~A. Pullin, Phys. Rev. Lett. {\bf 72},  3297  (1994).

\bibitem{Baker:2001sf}
J.~G. Baker, M. Campanelli, and C.~O. Lousto, Phys. Rev. D {\bf 65},  044001
  (2002).

\bibitem{Buonanno:1998gg}
A. Buonanno and T. Damour, Phys. Rev. D {\bf 59},  084006  (1999).

\bibitem{Buonanno:2000ef}
A. Buonanno and T. Damour, Phys. Rev. D {\bf 62},  064015  (2000).

\bibitem{Ajith:2007qp}
P. Ajith {\it et~al.}, Class. Quantum Grav. {\bf 24},  S689  (2007).

\bibitem{LIGODA}
B.~P. {Abbott} {\it et~al.}, Phys. Rev. X {\bf 6},  041014  (2016).

\bibitem{systematics}
B.~P. {Abbott} {\it et~al.}, Class. Quantum Grav. {\bf 34},  104002  (2017).

\bibitem{McWilliams:2010eq}
S.~T. McWilliams, B.~J. Kelly, and J.~G. Baker, Phys. Rev. D {\bf 82},  024014
  (2010).

\bibitem{Goebel}
C.~J. {Goebel}, Astrophys. J. {\bf 172},  L95  (1972).

\bibitem{Damour:2007xr}
T. Damour and A. Nagar, Phys. Rev. D {\bf 76},  064028  (2007).

\bibitem{Racine:2008kj}
E. Racine, A. Buonanno, and L.~E. Kidder, Phys. Rev. D {\bf 80},  044010
  (2009).

\bibitem{Nichols:2010qi}
D.~A. Nichols and Y. Chen, Phys. Rev. D {\bf 82},  104020  (2010).

\bibitem{Barausse:2011kb}
E. Barausse {\it et~al.}, Phys. Rev. D {\bf 85},  024046  (2012).

\bibitem{Ori:2000zn}
A. Ori and K.~S. Thorne, Phys. Rev. D {\bf 62},  124022  (2000).

\bibitem{Ames}
W.~L. {Ames} and K.~S. {Thorne}, Astrophys. J. {\bf 151},  659  (1968).

\bibitem{supp}
See Supplemental Material for additional comparisons of BOB and NR waveforms.

\bibitem{Ferrari}
V. {Ferrari} and B. {Mashhoon}, Phys. Rev. D {\bf 30},  295  (1984).

\bibitem{Yang}
H. {Yang} {\it et~al.}, Phys. Rev. D {\bf 86},  104006  (2012).

\bibitem{Misner73}
C.~W. Misner, K.~S. Thorne, and J.~A. Wheeler, {\em Gravitation} (W. H.
  Freeman, San Francisco, 1973).

\bibitem{Zerilli:1970se}
F.~J. Zerilli, Phys. Rev. Lett. {\bf 24},  737  (1970).

\bibitem{Poschl}
G. P{\"o}schl and E. Teller, Zeitschrift f{\"u}r Physik {\bf 83},  143  (1933).

\bibitem{Kelly}
B.~J. {Kelly} and J.~G. {Baker}, Phys. Rev. D {\bf 87},  084004  (2013).

\bibitem{Baker:2008mj}
J.~G. Baker {\it et~al.}, Phys. Rev. D {\bf 78},  044046  (2008).

\bibitem{Buonanno:2007sv}
A. Buonanno, L.~E. Kidder, and L. Lehner, Phys. Rev. D {\bf 77},  026004
  (2008).

\bibitem{Bardeen}
J.~M. Bardeen, W.~H. Press, and S.~A. Teuklolsky, Astrophys. J. {\bf 178},  347  (1972).

\bibitem{Kellyb}
B.~J. {Kelly} {\it et~al.}, Phys. Rev. D {\bf 84}, 084009 (2011).

\bibitem{Buonanno:2006ui}
A. Buonanno, G.~B. Cook, and F. Pretorius, Phys. Rev. D {\bf 75},  124018
  (2007).

\bibitem{Mroue}
A.~H. Mroue {\it et~al.}, Phys. Rev. Lett. {\bf 111},  241104  (2013).

\bibitem{catalog}
\url{http://www.black-holes.org/waveforms}.

\end{thebibliography}
\end{document}